\documentclass{JHEP}
\pdfoutput=1
\usepackage{amsmath}
\usepackage{graphicx}
\usepackage{multirow}

\usepackage{amsmath,amstext,amsfonts,amsbsy,amssymb,amscd,bbm,epsfig,lscape,cite}

\newcommand{\be}{\begin{equation}}
\newcommand{\ee}{\end{equation}}
\newcommand{\bea}{\begin{eqnarray}}
\newcommand{\eea}{\end{eqnarray}}

\title{Gain fractions of future neutrino oscillation facilities over T2K and NOvA}
\preprint{
	IFIC/13-07\\
	IFT-UAM/CSIC-13-019\\
	FTUAM-13-130 \\
	}

\author{M.~Blennow$^1$, P.~Coloma$^2$, A.~Donini$^{3,4}$,  E. Fern\'andez-Mart\'{\i}nez$^{3,5}$\\
(1) Department of Theoretical Physics, School of Engineering Sciences, \\
      KTH Royal Institute of Technology, AlbaNova University Center, \\
      106 91 Stockholm, Sweden \\
(2) Center for Neutrino Physics, Virginia Tech, Blacksburg, VA 24061, USA \\
(3) Instituto de F\'{\i}sica Te\'orica UAM/CSIC, \\
      Calle Nicol\'as Cabrera 13-15, E-28049 Madrid, Spain \\
(4)  Instituto de F\'{\i}sica Corpuscular, CSIC-Universitat de Val\`encia,\\
       Apartado de Correos 22085, E-46071 Valencia, Spain\\	
(5) Departamento de F\'{\i}sica Te\'orica UAM, \\
      Cantoblanco, E-28049 Madrid, Spain \\
}

\abstract{
We evaluate the probability of future neutrino oscillation facilities to discover leptonic CP violation and/or measure the neutrino mass hierarchy. 
We study how this probability is affected by positive or negative hints for these observables to be found at T2K and NO$\nu$A. We consider the following facilities: LBNE; 
T2HK; and the 10 GeV Neutrino Factory (NF10), and show how their discovery probabilities change with the running time of T2K and NO$\nu$A conditioned to their results. We find that, if after 15 years T2K and NO$\nu$A have not observed a $90 \%$ CL hint of CP violation, then LBNE and T2HK have less than a $10 \%$ chance of achieving a $5 \sigma$ discovery, whereas NF10 still has a $\sim 40\%$ chance to do so. Conversely, if T2K and NO$\nu$A have an early $90 \%$ CL hint in 5 years from now, T2HK has a rather large chance to achieve a $5 \sigma$ CP violation discovery ($75\%$ or $55 \%$, depending on whether the mass hierarchy is known or not). 
This is to be compared with the $90\%$ (30\%) probability that NF10 (LBNE) would have to observe the same signal at $5\sigma$. 
A hierarchy measurement at 5$\sigma$ is achievable at both LBNE and NF10 with more than 90\% probability, irrespectively of the outcome of T2K and NO$\nu$A.
We also find that if LBNE or a similar very long baseline super-beam is the only next generation facility to be built, then it is very useful to continue running T2K and NO$\nu$A (or at least T2K) beyond their original schedule in order to increase the CP violation discovery chances, given their complementarity.}

\begin{document}

\section{Introduction}
\label{sec:intro}

A coherent picture of leptonic mixing is gradually emerging from the experimental data, much as in the case of quark mixing: the three Standard Model lepton generations
mix via the unitary Pontecorvo-Maki-Nakagawa-Sakata\cite{Pontecorvo:1957cp,Pontecorvo:1957qd,Maki:1960ut,Maki:1962mu,Pontecorvo:1967fh} 
matrix $U_{\rm PMNS}$, the analogue of the CKM matrix $V_{\rm CKM}$  that governs hadronic mixing. This matrix can be parametrized in terms 
of three rotation angles, $\theta_{12}, \theta_{13}$ and $\theta_{23}$, that have been first measured using 
solar~\cite{Cleveland:1998nv,Kaether:2010ag,Abdurashitov:2009tn,Hosaka:2005um,Aharmim:2007nv, Aharmim:2005gt,Aharmim:2008kc,Bellini:2011rx}, reactor~\cite{An:2012eh,Ahn:2012nd,Abe:2011fz} and atmospheric~\cite{Wendell:2010md} neutrino data, respectively. 
Their values have been further constrained using long-baseline reactors~\cite{Gando:2010aa} and accelerator-based neutrino beams~\cite{Ahn:2002up,Ahn:2006zza,Adamson:2011qu,Abe:2011sj,Adamson:2012rm,Abe:2012gx}.
The most recent global fits of neutrino data give the following best-fits and 1$\sigma$ errors for the leptonic mixing angles and the neutrino squared mass differences~\cite{GonzalezGarcia:2012sz} (see also Refs.~\cite{Tortola:2012te,Fogli:2012ua}): 
\begin{equation}
\begin{array}{ll}
\left \{ 
\begin{array}{l}
\sin^2 \theta_{12} = 0.302^{\, + 0.013}_{\, - 0.012} \\
\\
\sin^2 \theta_{13} = 0.0227^{\, + 0.0023}_{\, - 0.0024} \\
 \\
\sin^2 \theta_{23} = 0.413^{\, + 0.037}_{\, - 0.025} \, / \,  0.594^{\, + 0.021}_{\, - 0.022} 
\end{array}
\right .
&
\, \left \{
\begin{array}{l}
\Delta m^2_{21} / 10^{-5 }= 7.50^{ \, + 0.18}_{\, - 0.19}  \,  {\rm eV}^2 \\
\\
\Delta m^2_{31}/10^{-3}  = 2.473^{\, + 0.070}_{\, - 0.067}  \,  {\rm eV}^2 \, ({\rm NH}) \\
\\
\Delta m^2_{32}/10^{-3}  =  - 2.427^{\, + 0.042}_{-0.065} \, {\rm eV}^2 \, ({\rm IH})
\end{array}
\right .
\end{array}
\end{equation}
where the twofold ambiguities related to allowed values of the $\theta_{23}$-octant and of the neutrino mass hierarchy are clearly shown.
These values strongly differ from the corresponding values obtained for the mixing angles in the CKM matrix~\cite{Beringer:1900zz}, 
when the latter is also parametrized in a similar way. It is well known that
the CKM matrix can be approximately parametrized as a small perturbation of the identity matrix, while the PMNS matrix rather displays entries which are all of similar order. The different patterns of quark and lepton mixing add to the puzzling {\em flavor problem}, {\em i.e.} the apparent arbitrariness of the Yukawa couplings of the Higgs boson with fermions belonging to different generations, 
that eventually induce fermion masses with large hierarchies after the spontaneous breaking of the electroweak symmetry.
The second important difference between the CKM matrix and the PMNS matrix
is that in the former case we know 
that the matrix is complex, {\em i.e.} that a non-vanishing CP-violating phase is needed in order to 
accurately reproduce the hadronic flavor mixing data. The quark CP-violating phase is $\delta_{\rm CKM} \sim \gamma =  \left ( 66 \pm 12 \right)^\circ$ (see, \emph{e.g.}, 
Refs.~\cite{DescotesGenon:2012rq} and \cite{Ricciardi:2013mca}).
A straightforward question is, therefore, if the leptonic mixing matrix is a complex matrix, as well as $V_{\rm CKM}$. 

The task of establishing whether $\delta_{\rm PMNS}$ is CP-violating (and possibly measuring it with a precision comparable to that of $\delta_{\rm CKM}$) 
is the main goal of the present generation of long baseline neutrino experiments. T2K~\cite{Itow:2001ee} is presently running, 
and the NO$\nu$A experiment~\cite{Ayres:2004js} is under construction and is expected to start data taking soon (in May 2013). 
Even if these experiments will be able to probe CP violation, their combined significance will remain below the $3\sigma$ evidence unless they run for an 
extended period of time, see for instance Refs.~\cite{Prakash:2012az,novawhitepaper}. 
Therefore, a next generation of long baseline neutrino experiments will most likely be needed if we want to establish leptonic CP violation at the $5 \sigma$ level. 
A second task that these experiments should address is the determination of the neutrino mass spectrum, {\em i.e.} if $\Delta m^2_{31} > 0$ (normal hierarchy) or
$\Delta m^2_{31} < 0$ (inverted hierarchy). However, due to the rather large value of $\theta_{13}$ recently discovered, the measurement of the neutrino mass hierarchy may be accessible to several alternative strategies to a long baseline neutrino experiment, such as atmospheric neutrino experiments~\cite{Blennow:2012gj,Akhmedov:2012ah,Ghosh:2012px,Agarwalla:2012uj,Franco:2013in} or a reactor experiment with a medium baseline~\cite{Petcov:2001sy,Choubey:2003qx,Zhan:2008id,Zhan:2009rs,Qian:2012xh,Qian:2012zn,Ciuffoli:2012bp,Ciuffoli:2012bs}.

In this paper we compare the proposed new facilities to be built in the following ten to twenty years based on their chances to establish a
CP-violating phase $\delta_{\rm PMNS}$ (from now on $\delta$, for simplicity) at a given CL and, secondarily, on their chances to pin down the correct neutrino mass hierarchy. 
For this purpose, we compute the probability that a given new facility has to discover a CP-violating $\delta$ (or a given mass ordering) at 5$\sigma$, 
conditioned to the presence or absence of a hint (at the 90\% CL for CP violation and at the 3$\sigma$ level for the mass hierarchy)\footnote{While higher CL can prove challenging to the statistically-limited T2K+NO$\nu$A, they will be able to provide 90\% hints over a large portion of the parameter space. Therefore, while we will require that the new experiment reaches the discovery limit, for T2K+NO$\nu$A only a hint at the 90\% CL will be required.} for the combination of T2K+NO$\nu$A. 
We define a condensed performance indicator, the \emph{gain fraction} $G$ (and $\bar G$), corresponding to the conditional probability that a next generation 
facility would reach the discovery limit over an observable, given that T2K+NO$\nu$A have (or have not) previously observed a hint of it. 
The {\em gain fraction} thus represents the ``gain'' in performance that the new facility can provide given the prior results from T2K and NO$\nu$A. This approach allows for a quantifiable {\em betting strategy} between the proposed new facilities, particularly important at this time of decision-making in neutrino oscillation experiments.

The paper is organized as follows. In Sect.~\ref{sec:method} we introduce the statistical approach and the terminology used in the rest of the paper before describing the technical details of the included facilities in Sect.~\ref{sec:setups}. We then present our results in Sect.~\ref{sec:results}, before we summarize and give our conclusions in Sect.~\ref{sec:concl}.

\section{The statistical method}
\label{sec:method}

We address the question of how much discovery potential can be gained by building a new facility given the prior information available at the time of 
its running from T2K and NO$\nu$A. If a significant hint for CP violation or a given mass hierarchy is already present, then a modest upgrade to increase the statistics should be able to provide the desired CL. Conversely, if no hint is found, a more demanding facility is needed to significantly increase our discovery potential. In other words, depending on the outcome of T2K and NO$\nu$A, can we quantify the additional gain a next generation facility would provide?

We will make use of some simple statistical tools. First of all, we compute the probability that a given facility combined with T2K+NO$\nu$A reaches a desired CL (outcome $A$) for a given observable (either discovery of CP violation or of the correct mass hierarchy) as a function of the true value of $\delta$, $P(A| \delta)$. In order to compute this probability we take the following steps: 
\begin{enumerate}
\item First, the expected event rates (including signal and backgrounds) are computed for each energy bin and oscillation channel of the facility under study, for an input value for $\delta$;
\item A Monte Carlo is then used to generate 1000 ``realizations'' of the experiment, by randomly generating numbers for the different energy bins and channels according to a Poisson distribution with mean value equal to the one computed in step 1. In order to do this, a new function has been added to the MonteCUBES free software~\cite{Blennow:2009pk} that produces a random distribution of events at a given experiment, provided its expected event rates;
\item For each realization of the experiment, we evaluate the CL with which CP violation or the correct mass hierarchy is favored.
Then, for a given value of $\delta$, we count the fraction of times the target CL was reached, that is, the fraction of times for which $A$ is fulfilled, thus estimating the probability 
$P(A| \delta)$.
\end{enumerate}    
The probability $P(A|\delta)$ is the starting point of our statistical analysis. In particular, instead of computing the expected achievable CL for a given observable
({\em e.g.} the capability to observe a CP-violating signal) at a given facility, we rather fix a \emph{target} CL (typically $90\%$, $3\sigma$ 
or $5\sigma$) and ask what is the probability with which the facility will reach the target.
This procedure is slightly different from what is commonly done in the literature, where the most widely used performance indicator is rather the discovery potential 
for a given observable, defined as the CL that a facility is expected to provide for that observable as a function of $\delta$. In this case,
the expected number of events for the facility is computed for each value of $\delta$ and the CL with which CP conservation or the wrong mass hierarchy would be disfavored for that value of $\delta$ is evaluated. It has been shown (see Ref.~\cite{Schwetz:2006md}) that the probability $P(A| \delta)$ of the facility actually reaching 
that CL is around 50\%. Indeed, upward or downward random fluctuations will increase or decrease the final CL in a particular realization of the experiment.

To address the question of how much we can improve on the results from T2K and NO$\nu$A we will also make use of the {\em joint} probability $P(A,B| \delta)$, 
representing the probability that the new facility combined with T2K+NO$\nu$A reaches a given CL (outcome $A$) and that the combination of T2K+NO$\nu$A alone also observed a signal at some CL for the same observable (outcome $B$)\footnote{
We will also let $\bar B$ represent the complement of $B$, \emph{i.e.}, that T2K+NO$\nu$A do not reach a given CL. We then have $P(A,B| \delta) + P(A,\bar{B}| \delta) = P(A| \delta)$.}, for a given $\delta$. It should be noted that the CL required for outcomes $A$ and $B$ do not need to be the same. The computation of the joint probability 
is done following similar steps to 1-3. We first compute the expected event rates 
for both T2K and NO$\nu$A and the next generation facility and generate 1000 different realizations of them. Then, for each realization, 
we record the CL with which T2K and NO$\nu$A, on their own as well as in combination with the new facility, are able to disfavor CP conservation 
or the wrong mass hierarchy. Finally, we count how many times T2K and NO$\nu$A were (un)able to reach the desired CL while the combination with the new facility reached its desired CL. This procedure is repeated for all values of $\delta$.

We are interested, in particular, in the joint probability that T2K+NO$\nu$A find (or not) a hint at the 90\% CL for a given value of $\delta$ ($B$ or $\bar{B}$, respectively) and that their combination with a new facility reaches 5$\sigma$ for the same $\delta$ and observable ($A$). 
In the case of the mass hierarchy we will raise the CL of $B$ to the $3\sigma$ level, since T2K+NO$\nu$A will most likely have a 90\% hint.
The joint probability $P(A,\bar{B}|\delta)$ defined this way is a measure of the complementarity between T2K+NO$\nu$A and the new facility for that particular value of $\delta$. Indeed, if $P(A,\bar{B}|\delta)$ is high it means that there is a high probability that T2K+NO$\nu$A will \emph{not} find a hint at that value of $\delta$ while the probability of making a discovery of their combination with the new facility is also high. Clearly, for facilities whose best chances to measure $\delta$ take place for the same region of parameter space as for the combination of 
T2K and NO$\nu$A, the joint probability $P(A,\bar{B}|\delta)$ will be small
, since for favorable values of $\delta$ a positive result at the 
90\% CL would have been likely at T2K+NO$\nu$A. On the other hand, for a facility whose best sensitivity to $\delta$ is not totally overlapping 
with that of T2K+NO$\nu$A, we expect a rather high joint probability in the regions where they do not overlap. 
In order to simplify for the reader to recollect our definitions of $A$ and $B$, these can be found in Tab.~\ref{tab:ABdef}.
\begin{table}
\begin{center}
\begin{tabular}{|l|p{12cm}|}
\hline
{\bf Event} & {\bf Definition} \\
\hline
$A$ & The combination of T2K+NO$\nu$A with a future facility provides a discovery of CP violation or the mass hierarchy at $5\sigma$. \\
\hline
$B$ & T2K+NO$\nu$A provide a signal for CP violation (or mass hierarchy) at 90\% CL (3$\sigma$). \\
\hline
$\bar B$ & T2K+NO$\nu$A do not provide a hint. The complement of $B$. \\
\hline
\end{tabular}
\caption{ \it The definitions of the events $A$, $B$, and $\bar B$. Cited for back reference. \label{tab:ABdef}}
\end{center}
\end{table}

A very interesting (and condensed) quantity to parametrize the margin for improvement over the results of T2K+NO$\nu$A by a new facility is 
the {\em conditional probability} $P(A|B) = P(A,B)/P(B)$. This is the probability of achieving a $5\sigma$ discovery at the new facility 
provided a 90\% CL or $3\sigma$ signal has been previously obtained at
T2K+NO$\nu$A. For brevity, we will call this quantity the {\em gain fraction} $G$. Similarly, we can define the gain fraction 
assuming the opposite outcome: $\bar{G} = P(A|\bar{B})$,
\emph{i.e.}, the probability of achieving a $5\sigma$ discovery at the new facility provided no 90\% CL or $3\sigma$ signal has been previously seen at T2K+NO$\nu$A. 
Notice that we have defined the gain fractions in a way that does not depend on $\delta$. Indeed, the outcome from T2K+NO$\nu$A 
will provide a probability distribution for the possible values 
of $\delta$, $P(\delta|B)$, that can be used as \emph{prior} information for the measurements of the new facility. Integration can be performed over $\delta$ as follows\footnote{Using that $P(\delta |B) = P(B|\delta) \times P(\delta)/P(B)$ according to Bayes' theorem. We also assume a flat distribution for $\delta$, $P(\delta) = 1/(2\pi)$, prior to the measurement from T2K+NO$\nu$A.}:
\begin{equation}
G = P(A|B) = \int P(A|B,\delta) P(\delta|B) d \delta   = \frac{\int P(A,B|\delta) d \delta}{\int P(B|\delta) d \delta} = \frac{P(A,B)}{P(B)} \, .
\end{equation}
We can see that there is a very close relation between the joint probability $P(A,B|\delta)$ and the gain fractions. Thus, the joint probability is not only useful as a 
measure of the complementarity of a new facility with respect to T2K+NO$\nu$A, but can also help to understand the behaviour of the gain fraction. 
For this reason, example plots of the joint probability for the CP and mass hierarchy discovery as a function of $\delta$ will also be shown in Sect.~\ref{sec:results}. 
We will also show the dependence of the gain fractions $G$ and $\bar{G}$ on the running time of T2K+NO$\nu$A. From the above equation, it is clear that the gain fraction can also be obtained in a graphical way: it is simply the area under the curve for $P(A,B|\delta)$ divided by the area under $P(B|\delta)$, thus representing the fraction of parameter space remaining after T2K+NO$\nu$A that is probed by the new facility. For zero running time of T2K+NO$\nu$A the gain fraction is, therefore, 
intimately related to the CP-fraction that is usually discussed in the literature.

Finally, it should be noted that $B$ and $\bar{B}$ are oversimplifications of the results from T2K+NO$\nu$A. Indeed, the outcome from T2K+NO$\nu$A 
will not be limited to the presence or absence of a hint, but will rather be some measured event distributions that would translate in a given 
probability distribution for $\delta$ when trying to fit different values to data, $P(\delta|{\rm data})$. 
It is this probability distribution which should be used as condition $B$ for the prior when computing the gain fraction. 
However, lacking the results from T2K+NO$\nu$A we decided to adopt two representative scenarios to present our results: the presence or absence of a 90\% CL or $3\sigma$ signal.

\section{Setups}
\label{sec:setups}

\begin{table}[t!]
\begin{center}{
\renewcommand{\arraystretch}{1.5}
\begin{tabular}{|l|clclcccc|}
\hline
{\bf Setup} & {\bf $E^\textrm{peak}_\nu$} & {\bf $L$} & {\bf OA} & {\bf Detector} & {\bf kt} & {\bf MW} & {\bf Decays/yr} & {\bf ($t_\nu$,$t_{\bar\nu}$)}  \\
\hline
T2K & 0.6 & 295 & $2.5^\circ$ & WC & 22.5  & 0.2-0.7 & -- & Var. \\  
NO$\nu$A & 2 & 810  & $0.8^\circ$ & TASD  & 3-14  & 0-0.7 & -- & Var.  \\ 
\hline
NF10 & 6  & 2000 & -- & MIND & 100 & --   & 7$\times10^{20}$ & (10,10)  \\ 
LBNE & 3.0 & 1290 & -- & LAr  & 10-33 & 0.8  & -- & (5,5) \\ 
T2HK & 0.6 & 295  & $2.5^\circ$  & WC   & 560 & 1.66 & -- & (1.5,3.5)  \\ \hline
\end{tabular}}
\end{center}
\caption{ \it \label{tab:setups} Main features of the setups considered in this
work. From left to right, the columns list the energy at which the number of neutrino events peak (in GeV), the length of the baseline (in km), the off-axis angle in degrees, the detector technology (Magnetized Iron Neutrino Detector, Liquid Argon, Water \v{C}erenkov or Totally Active Scintillator Detector), the fiducial mass of the detector, the beam power or the number of useful decays per year, and the number of years that the experiment is assumed to run per polarity. The label ``Var.'' indicates that the running times for T2K and NO$\nu$A are variable, see text for details. 
 }
\end{table}

\begin{figure}
\begin{center}
\begin{tabular}{cc}
  \includegraphics[scale=0.55]{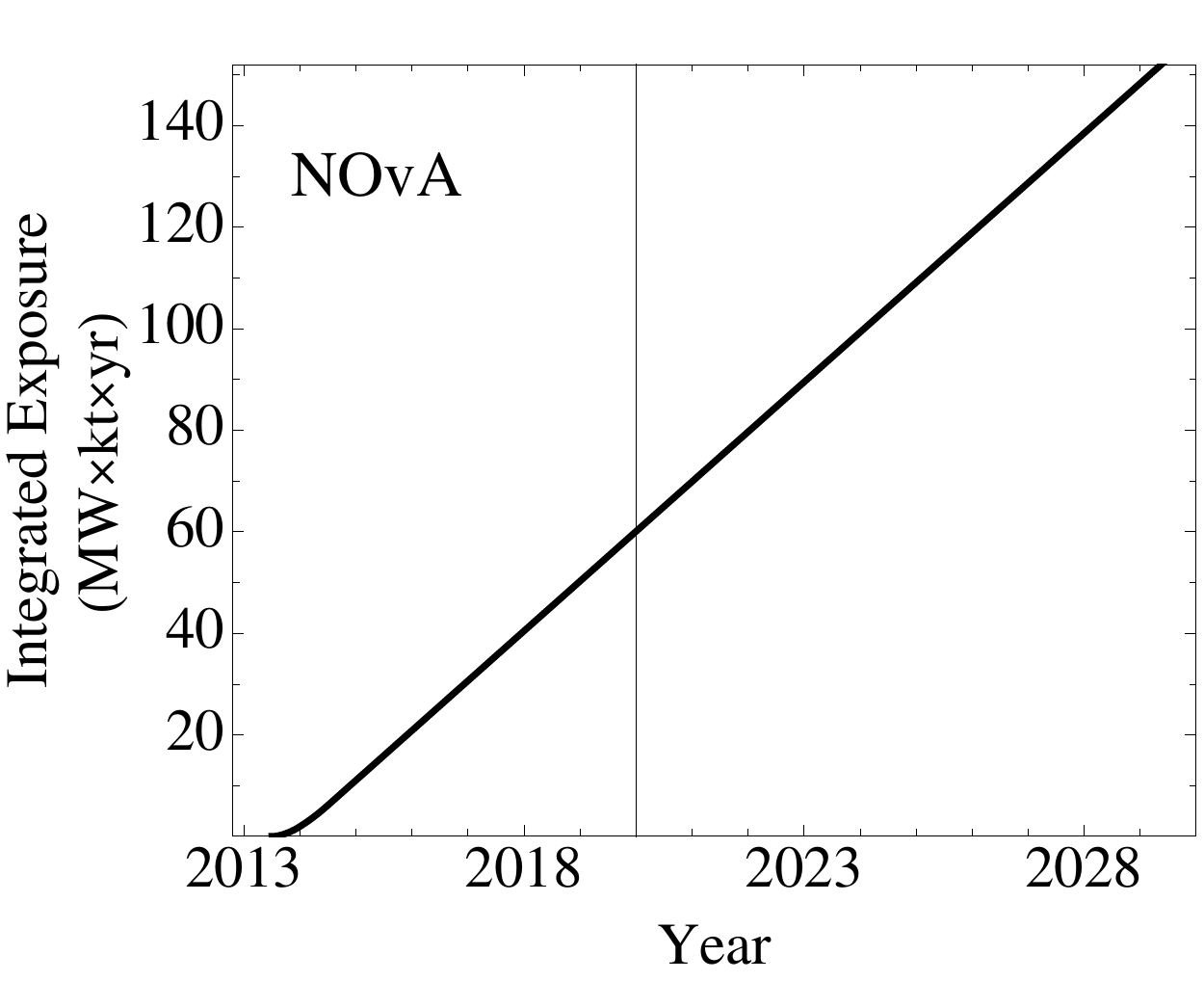} &
  \includegraphics[scale=0.55]{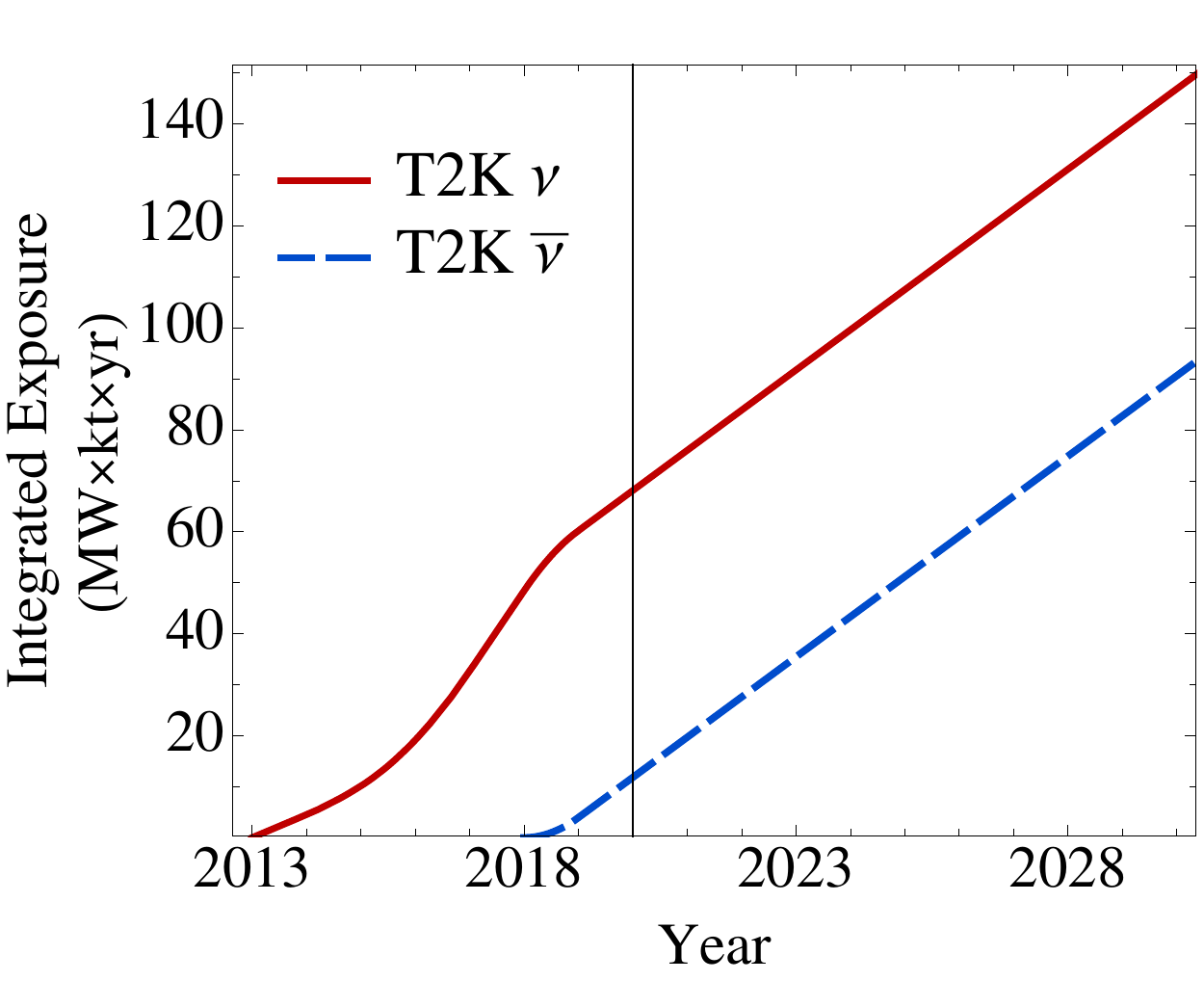}
\end{tabular}
\end{center}
\caption{\label{fig:expo} \it Integrated exposure as a function of time for NO$\nu$A (left panel) and T2K (right panel). For NO$\nu$A, the running time is split equally between neutrino and antineutrino modes. The shown integrated exposure corresponds to the total between the two polarities. For T2K, the exposure for each polarity is assumed to change differently with time, as seen in the right hand panel. The integrated exposure in this case corresponds to each polarity separately. 
The vertical lines correspond to year 2020.
}
\end{figure}

The performance of new facilities will be compared to (and combined with) those attainable from the combination of T2K and NO$\nu$A, assuming variable running times. The main features for these two experiments are summarized in Tab.~\ref{tab:setups}. Notice that the beam power of T2K has not reached its goal yet and it is not expected to do so in the next couple of years. In the case of NO$\nu$A, both the fiducial volume of the detector as well as the beam power are expected to increase progressively within the first few months of data taking of the experiment, which is expected to start in May 2013~\cite{Patterson:2012zs}. 
Several assumptions have been made in this work in order to perform as realistic as possible an estimate of the statistics achievable at each experiment for a given date. Fig.~\ref{fig:expo} shows our assumptions for the integrated exposure for T2K and NO$\nu$A as a function of time. The integrated NO$\nu$A exposure has been extracted from Ref.~\cite{Patterson:2012zs}. For T2K we have followed Ref.~\cite{T2Kexpo}, assuming that the experiment starts in 2013 at 200 kW, goes up to 300 kW in 2014, and finally reaches 700 kW in 2017. In addition, we have assumed that during the first 5 years it is only run in neutrino mode, but that after that date the running is equally split between the two polarities. A vertical line corresponding to year 2020 has been added to Fig.~\ref{fig:expo} for reference.

The simulation of the T2K and NO$\nu$A experiments is largely based on Ref.~\cite{Huber:2009cw}. The T2K simulation is exactly the same, except for the integrated exposure which has been varied according to Fig.~\ref{fig:expo}. The situation is a bit different for NO$\nu$A. In view of the large value of $\theta_{13}$ the NO$\nu$A collaboration have recently modified their analysis techniques~\cite{Patterson:2012zs}. In particular, some of the cuts have been relaxed so that larger efficiencies are now expected, at the price of larger backgrounds. Therefore, we have used the same signal and background rejection efficiencies as in Ref.~\cite{Agarwalla:2012bv}, which are in agreement with those in Ref.~\cite{Patterson:2012zs}. Finally, we are using matrices to migrate NC backgrounds to low energies, unlike in Ref.~\cite{Huber:2009cw} where migration matrices were not available for this purpose.

Alongside with T2K and NO$\nu$A,  we consider in this paper a few representative facilities to analyse under the new performance indicators described in 
Sect.~\ref{sec:method}. We will concentrate in particular on three of the most discussed setups for the next generation of neutrino oscillation facilities: 
two super-beams, T2HK and LBNE, 
and the ``low energy" \cite{Geer:2007kn,Ankenbrandt:2009zza,FernandezMartinez:2010zza,Dighe:2011pa,Ballett:2012rz} Neutrino Factory (NF10), 
which is considered as the ultimate neutrino oscillation facility but also a longer term project.

The technical details of the simulations for these three facilities are summarized in Tab.~\ref{tab:setups}: 
\begin{enumerate}
\item {\bf T2HK} \\
This setup uses the same baseline ($L = 295$ km) and accelerator as T2K (albeit, with a higher power of 1.66 MW) aiming at a new
detector, Hyper-KamiokaNDE (HK), with a fiducial volume of 560 kton of water~\cite{Abe:2011ts}. To compensate for the different 
$\nu N$ and $\bar \nu N$ CC cross-sections, this setup is foreseen to run for 1.5 years with $\pi^+$ and for 3.5 years with $\pi^-$. 
The collaboration is also exploring the possibility to run the experiment for 10 years (3 years with $\pi^+$ and 7 years for $\pi^-$) but with a 
beam power of 750 kW (see, for instance, Refs.~\cite{HKnuturn,openHKmeeting}). 
We have checked that the results using this configuration are very similar, and therefore will only show those corresponding to the configuration 
described in the HK letter of intent, Ref.~\cite{Abe:2011ts}.
\item {\bf LBNE} \\
The proposal consists of a new accelerator to be built at FermiLab with a design power of 0.7 MW, aiming at a distant, on-axis, 33 kton Liquid Argon (LAr) detector 
(in a first phase, only 10 kton of LAr are considered) located at $L = 1290$ km from the source\footnote{European alternatives to these two facilities with somewhat similar physics performances have also been proposed. The SPL~\cite{Campagne:2006yx,Coloma:2011pg} or ESS proposals~\cite{Baussan:2012cw} present similar characteristics and performance to T2HK, while LBNO~\cite{LBNO}, with a smaller LAr detector, longer baseline and higher energy, has a similar performance to LBNE.}. Five years of data taking for $\pi^+$ and $\pi^-$ 
are considered. The simulation details have been implemented according to Ref.~\cite{Akiri:2011dv}. 
\item {\bf NF10} \\
After the measurement of a large $\theta_{13}$, a re-evaluation of the Neutrino Factory design~\cite{Choubey:2011zzq} has identified the optimal parent muon energy for this
facility at 10 GeV (see also Ref.~\cite{Agarwalla:2010hk}). The new design includes a 100 kton magnetized iron neutrino detector (MIND, \cite{Bayes:2012ex}), located at approximately $L = 2000$ km from the source. Both $\mu^+$ and $\mu^-$ would run at the same time inside the decay ring for a total of 10 years. The detector simulation details have been implemented according to Ref.~\cite{Bayes:2012ex}. Migration matrices for both signal and backgrounds have been used~\cite{bayes}, and the so-called $\tau$-contamination~\cite{Indumathi:2009hg,Donini:2010xk,Coloma:2011zza,Dutta:2011mc} has been included in the analysis for both appearance and disappearance signals.
\end{enumerate}

Systematic errors are expected to have a large impact on the results for long baseline experiments, given the large value of $\theta_{13}$. In Ref.~\cite{Coloma:2012ji} a detailed analysis of systematics and their impact on long baseline oscillation experiments was performed. It was shown that the effect on LBNE as well as the NF10 setups is small\footnote{Note that no uncertainties on the shape of the neutrino flux or cross section were considered in Ref.~\cite{Coloma:2012ji}, though. If this is not the case the situation may be very different, see for instance Refs.~\cite{Martini:2012fa,Nieves:2012yz,Lalakulich:2012hs}.} if a fully correlated analysis is performed. In the case of T2HK, however, the results present large variations depending on the particular values for the systematic uncertainties and, more importantly, on the particular implementation used. In the present work, systematic uncertainties have been implemented as overall normalization errors, which are uncorrelated between signal and backgrounds but correlated among all backgrounds contributing to the same channel. Even if this does not correspond to the real physics case, it was shown in Ref.~\cite{Coloma:2012ji} that an appropriate choice of the normalization errors would yield similar results to those obtained with a more sophisticated systematics implementation. Thus, we have chosen the following normalization errors for the signal and backgrounds in each experiment: 1.5\% and 10\% for the NF10, 3\% in both cases for LBNE, and 5\% and 10\% for T2HK  and NO$\nu$A. For T2K, on the other hand, 2\% and 5\% normalization errors have been used for the signal and background event rates, plus a calibration (tilt) error of 1\% and 5\%, respectively, following Ref.~\cite{Huber:2009cw}.

\section{Results}
\label{sec:results}

In this section we apply the statistical method described in Sect.~\ref{sec:method} to compare the different facilities defined in Sect.~\ref{sec:setups}, under the 
hypotheses that the combination of the data taken by T2K and NO$\nu$A, after a given number of years of running time, either does or does not show a hint of a CP-violating phase $\delta$ (or of a given hierarchy) at the 90\% CL (3$\sigma$). We restrict to this CL due to the statistical limitations of both experiments. 

\subsection{CP violation}
\label{sec:resCP}

\begin{figure}[htb!]
\begin{flushleft}
	\includegraphics[width=17cm]{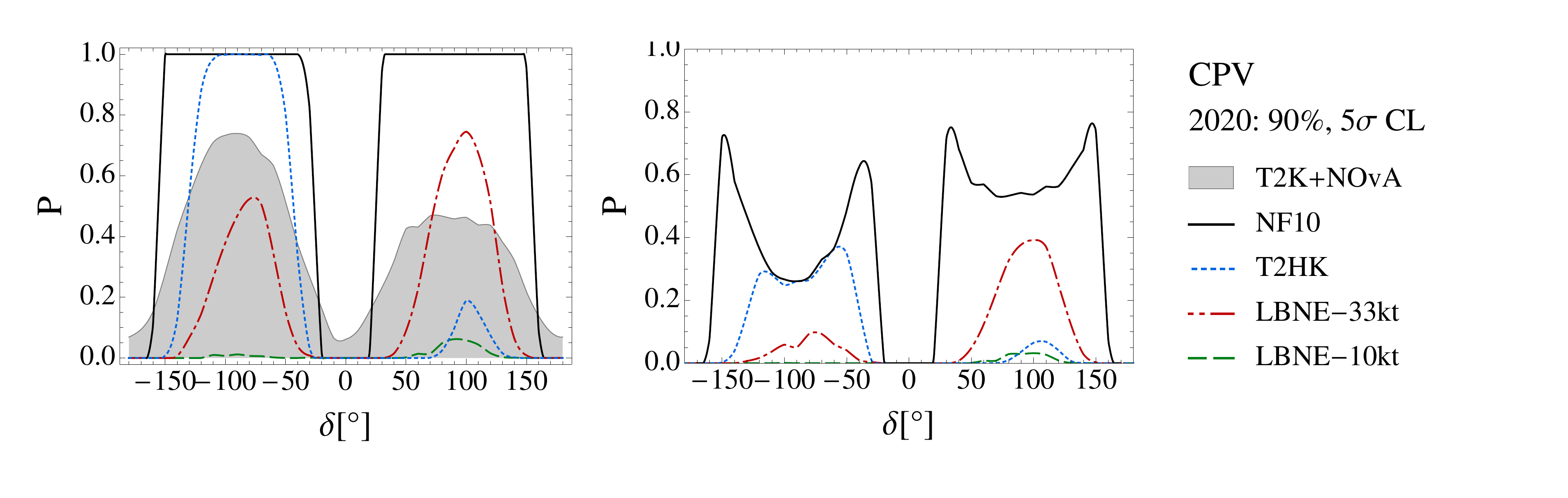}
\caption{\it 
Left panel: probability of finding CP violation (CPV) at the 90\% CL for the combination T2K+NO$\nu$A (gray-shaded area); 
probability of finding CP violation at 5$\sigma$ for the new facilities in combination with the results obtained at T2K+NO$\nu$A (lines). 
Right panel: joint probability that a new facility in combination with T2K+NO$\nu$A observes CP violation at 5$\sigma$ and that T2K+NO$\nu$A alone have not seen a hint of it at 90\% CL for the same value of $\delta$. In all cases, it is assumed that T2K+NO$\nu$A will stop data taking on 2020.
 }
\label{fig:202090CL5sigdelta}
\end{flushleft}
\end{figure}

Fig.~\ref{fig:202090CL5sigdelta} shows the results for T2K+NO$\nu$A assuming that they will take data until 2020 ({\em i.e.}, close to the initially foreseen running time). 
The probability of finding CP violation at 90\% CL for the combination of T2K+NO$\nu$A is represented in Fig.~\ref{fig:202090CL5sigdelta} (left) by the gray-shaded area. 
In the same panel, the lines show the probability of finding CP violation (\textit{i.e.}, to distinguish a CP-violating phase $\delta$ from 0 and $\pi$) at 5$\sigma$ for the different new facilities described in Sect.~\ref{sec:setups}, in combination 
with T2K+NO$\nu$A data. Solid black lines correspond to NF10; dotted blue lines to T2HK; dot-dashed red lines to LBNE-33 kton; 
and dashed green lines to LBNE-10 kton. 
As explained in Sect.~\ref{sec:method}, this probability is obtained by simulating different realizations of a given experiment 1000 times for each input value of $\delta$ and checking whether in that particular realization the facility is able to exclude CP-conserving values of $\delta$ at the desired CL.
For T2K+NO$\nu$A and for T2HK, the probability of finding CP violation is strongly asymmetric, reaching higher probabilities for negative $\delta$. This is a consequence of the sign degeneracy that leads to allowed CP-conserving regions for positive $\delta$ when measuring $\nu_\mu \rightarrow \nu_e$ oscillations~\cite{Minakata:2001qm,Donini:2003vz,Huber:2002mx}.

To explain our results let us consider now, for example, the black solid line in Fig.~\ref{fig:202090CL5sigdelta} (left), 
that represents the probability of finding CP violation at 5$\sigma$ (condition $A$)
for\footnote{We have checked that the combination of the NF10 data with T2K+NO$\nu$A is irrelevant, due to the low statistics of the latter 
with respect to the NF10.} the NF10, $P(A|\delta)$.
Two flat regions can be seen for $\delta \in [-150^\circ,-30^\circ]$ and $[30^\circ,150^\circ]$, for which the probability to measure a CP-violating phase at 5$\sigma$ at the NF10 is $P (A|\delta) \sim 1$. 
Depending on $\delta$, it is more or less likely that T2K+NO$\nu$A will have already found a 90\% CL hint by themselves.
For maximally CP-violating values of $\delta$, in particular, there is a fair chance that they will have found a hint by 2020 and, thus, condition $\bar B$
{\em will not} be satisfied. This is reflected in that the black solid line in Fig.~\ref{fig:202090CL5sigdelta} (right), which represents the {\em joint probability} 
$P (A,\bar B |\delta)$, is reduced for $|\delta| = 90^\circ$. For $\delta = -90^\circ$, where T2K+NO$\nu$A have a peak in the probability of finding CP 
violation at the 90\% CL, we see that $P(A|-90^\circ) \sim 1$ whereas $P(A,\bar B|-90^\circ) \sim 0.3$.
On the borders of the region of maximal probability of finding CP violation for T2K+NO$\nu$A we see high spikes for the NF10 \emph{joint probability}. These appear because the flat region that we observe in Fig.~\ref{fig:202090CL5sigdelta} (left) for NF10 is much broader than the region where the probability of finding CP violation for T2K+NO$\nu$A (the gray-shaded area) is maximal. This shows that, in those spikes, NF10 can probe for CP violation with very high significance, while T2K+NO$\nu$A have a hard time finding a hint. Thus, the gain factor $\bar{G}$ for NF10 will receive significant contributions from those areas of the parameter space.

\begin{figure}[htb!]
\begin{flushright}
	\includegraphics[width=13cm]{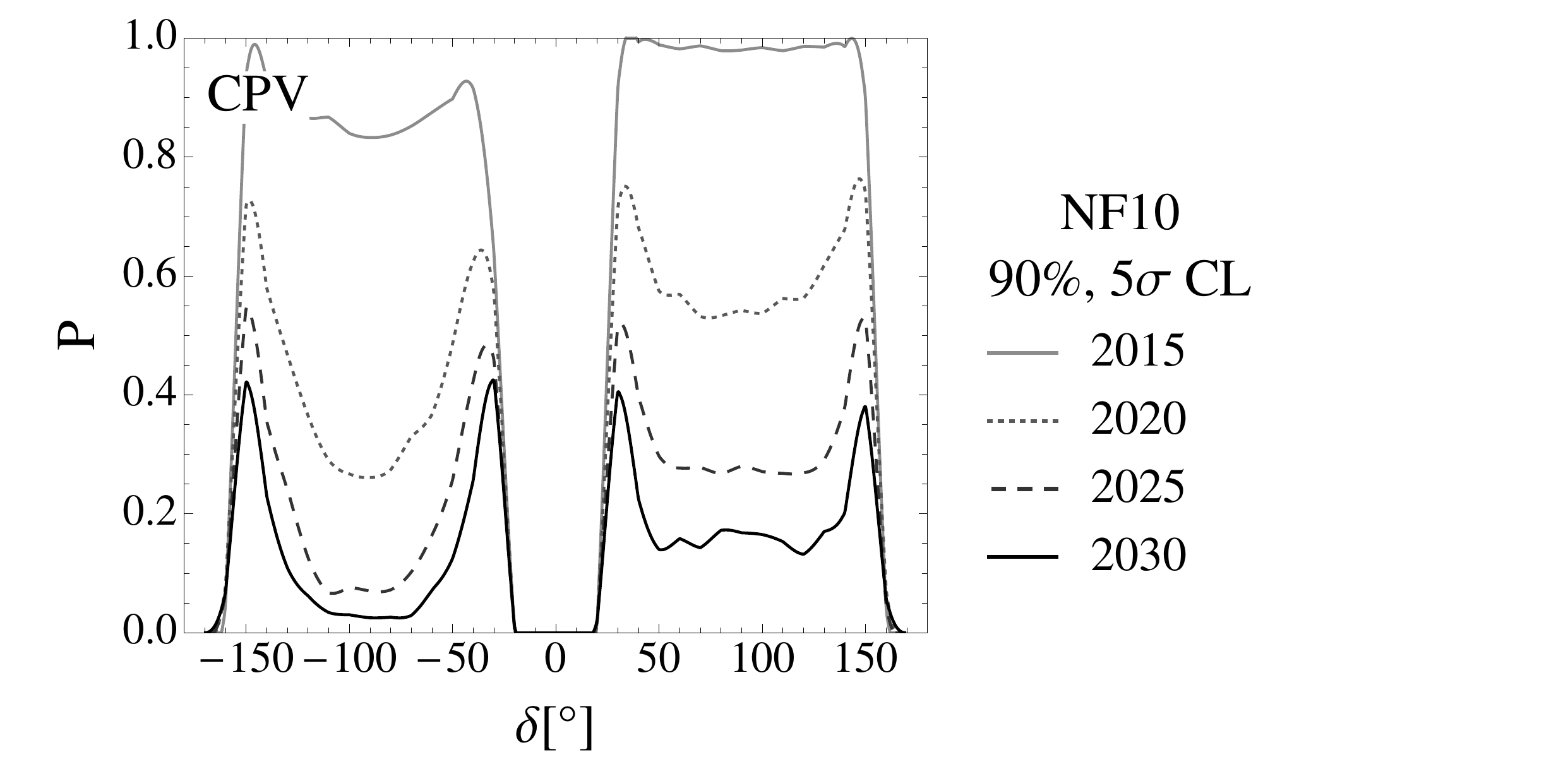}
\caption{\it  Joint probability that NF10 observes CP violation at $5 \sigma$ for a fixed value of $\delta$ and that T2K+NO$\nu$A have not been able to see a hint of it at the 90\% CL for the same value of $\delta$. Results are shown assuming that T2K+NO$\nu$A stop data taking in 2015, 2020, 2025 or 2030, as indicated in the legend.} 
\label{fig:deterioration}
\end{flushright}
\end{figure}

It is interesting to observe how the {\em joint probability} $P(A,\bar B|\delta)$ is affected as the running time of T2K+NO$\nu$A is increased. This is illustrated 
in Fig.~\ref{fig:deterioration}, where we show the {\em joint probability} that NF10 discovers CP violation at $5 \sigma$ (condition $A$) 
and T2K+NO$\nu$A have not seen a hint of it at the $90 \%$ CL if they stop data taking on 2015, 2020, 2025 or 2030 (condition $\bar B$). It can be clearly seen that the probability
that both conditions are fulfilled for maximal CP-violating values of $\delta$ ($\delta \sim \pm 90^\circ$) rapidly decreases with increasing T2K+NO$\nu$A exposures,
whereas spikes at $\delta \sim \pm 30^\circ, \pm 150^\circ$ are only mildly affected (showing that, for NF10, some areas of the parameter space remain promising even after a negative result from T2K+NO$\nu$A). The reason for this is that, as the data taking period increases for T2K+NO$\nu$A, it is more and more unlikely that they do not observe a hint for CP violation at the 90\% CL if $\delta$ is maximally CP-violating.

\begin{figure}[htb!]
\begin{flushleft}
	\includegraphics[width=17cm]{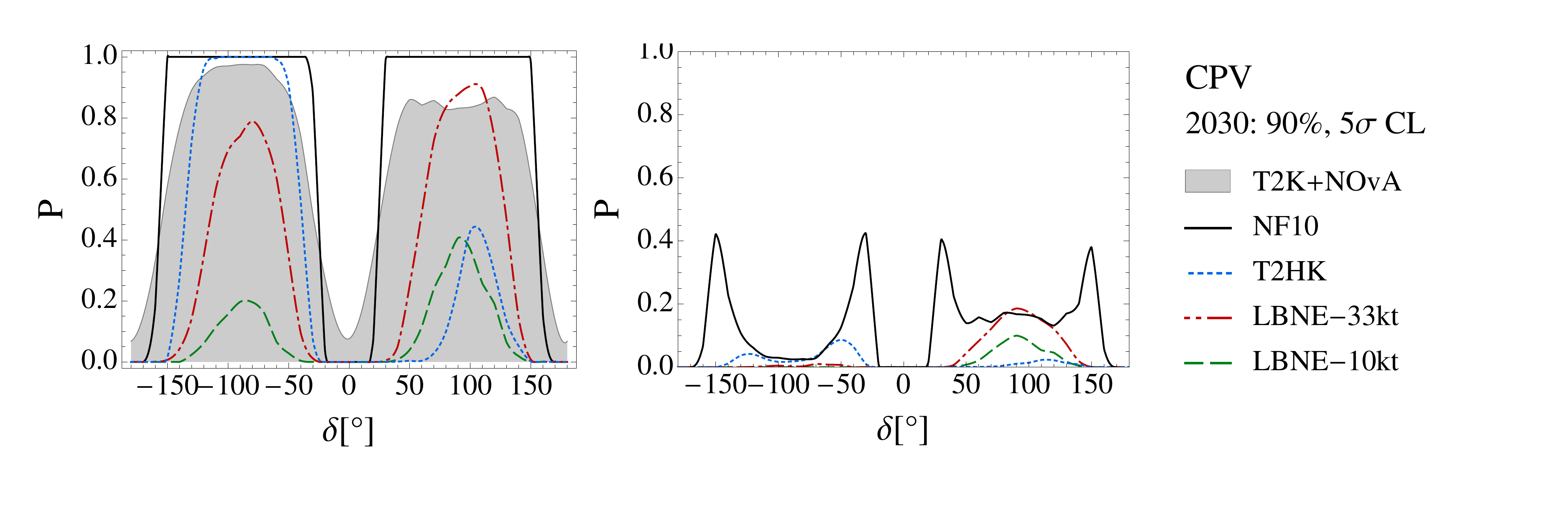}
\caption{\it 
Same as in Fig.~\ref{fig:202090CL5sigdelta} assuming that T2K+NO$\nu$A will run until 2030. }
\label{fig:203090CL5sigdelta}
\end{flushleft}
\end{figure}

The same effect is illustrated in Fig.~\ref{fig:203090CL5sigdelta} for all facilities, which shows the same probabilities as in Fig.~\ref{fig:202090CL5sigdelta} but assuming that T2K+NO$\nu$A run for 10 additional years, until 2030.
As it can be seen, the probability of not finding a positive $90 \%$ CL hint for $\delta \sim \pm 90^\circ$ if T2K+NO$\nu$A keep running until 2030 is very small. This in turns implies that, provided a negative result from T2K+NO$\nu$A, CP-violating values for $\delta$ are much less likely and the new facility will have smaller chances of finding CP violation. The only facility that retains a non-negligible joint probability after this exposure is NF10. This fact is better captured by the {\em gain fractions} $G$ and $\bar G$ defined in Sect.~\ref{sec:method}, which correspond to the conditional probability 
that the new facility will detect CP violation at 5$\sigma$, provided that T2K+NO$\nu$A have previously seen ($G$) or have not seen ($\bar G$) 
a hint at the 90\% CL. 

\begin{figure}[htb!]
\begin{flushleft}
\begin{tabular}{cc}
	\includegraphics[width=7cm]{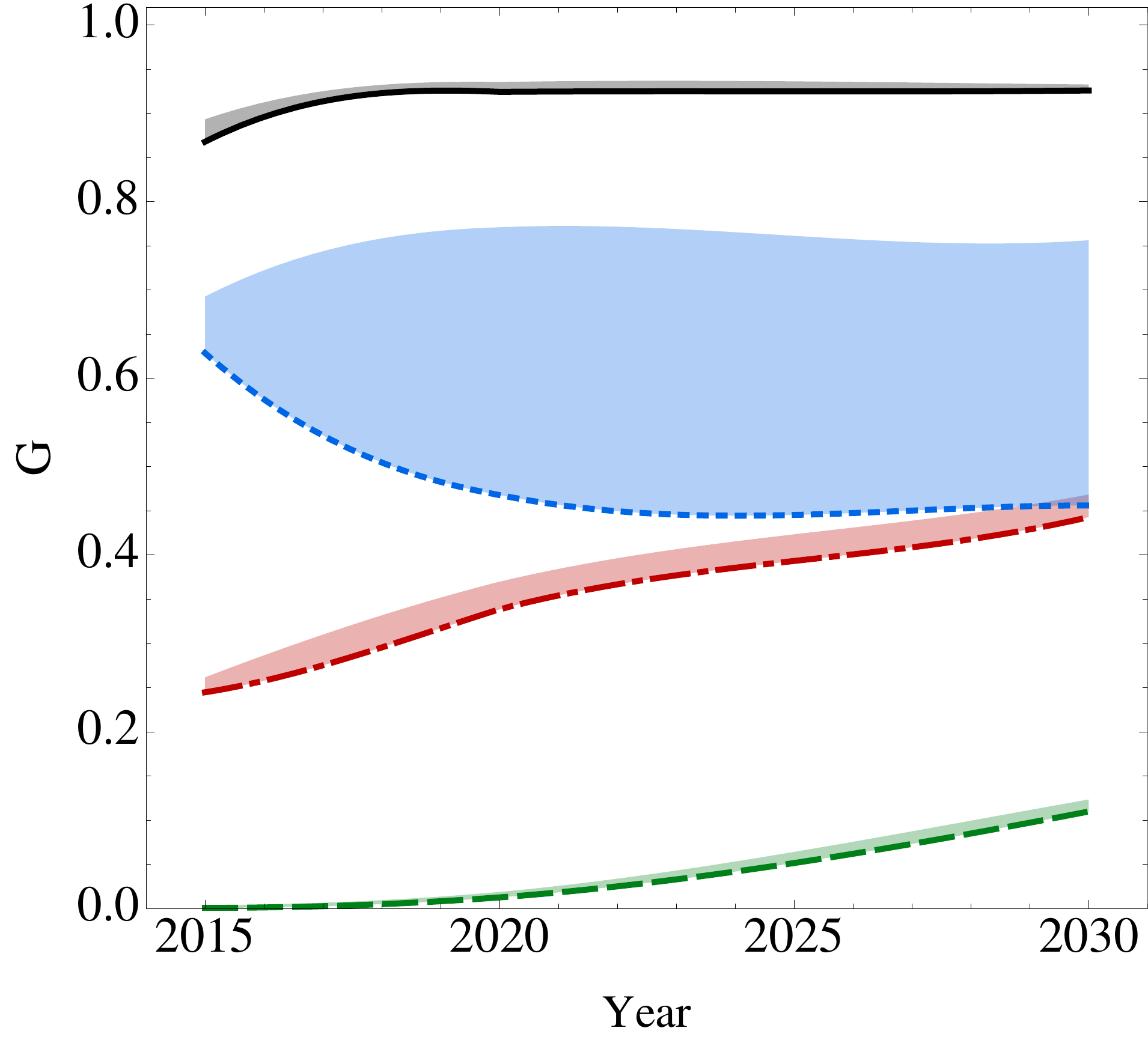} &
	\includegraphics[width=7cm]{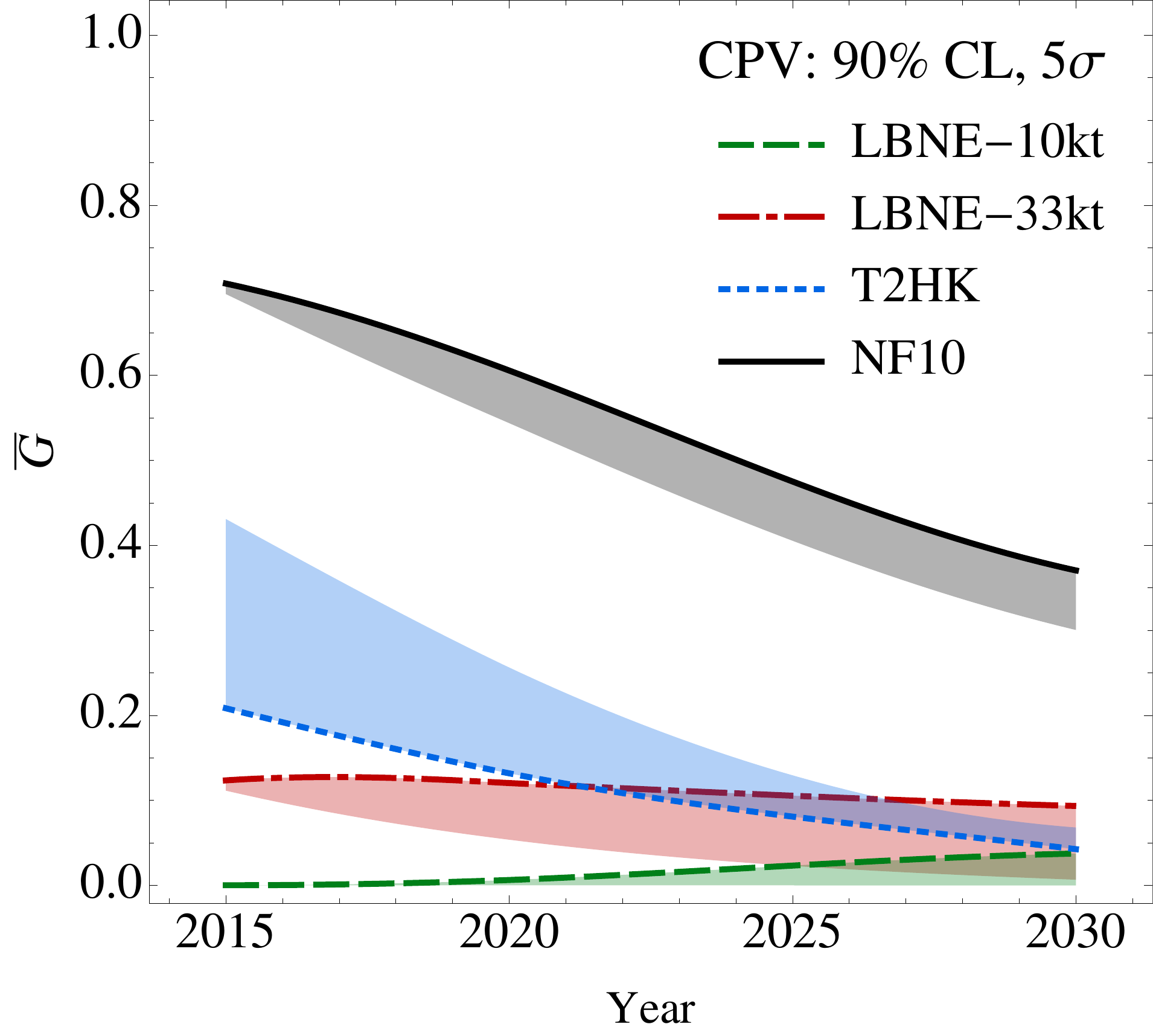}
\end{tabular}
\caption{\it The gain fractions $G$ (left panel) and $\bar G$ (right panel) for CP violation defined as in Sect.~\ref{sec:method}, {\em i.e.} the probability that the new facility will detect a CP-violating signal at 5$\sigma$ provided that T2K+NO$\nu$A have seen or not a hint of it at 90\%CL irrespectively of the value of $\delta$. In both cases results are shown as a function of  the year when T2K+NO$\nu$A stop taking data.
Bands show how the gain fractions change if we assume no previous knowledge of the mass hierarchy (thick edges) or that the mass hierarchy
is known by the time the new facility analyses its data (thin edges).}
\label{fig:gainfractions905sigdelta}
\end{flushleft}
\end{figure}

We show in Fig.~\ref{fig:gainfractions905sigdelta} bands for $G$ (left panel) and $\bar G$ (right panel) as a function of the year when T2K+NO$\nu$A stop data taking.
The lower and upper limits of each band correspond to two possible scenarios: (1) the neutrino mass hierarchy is not known at the time when the data of T2K+NO$\nu$A are analyzed, and (2) the neutrino mass hierarchy is known at 5$\sigma$ at that time, assuming that other experiments have been able to measure it. The gain fractions obtained in case 1 
are depicted by thick lines, whereas those corresponding to case 2 are shown as just the end of each band. As shown in the figure, $G$ always increases when the additional information about the mass hierarchy is added, while the situation is the opposite for $\bar G$ (with the exception of T2HK).
In case the mass hierarchy is measured during a certain year, it suffices to consider the band edge corresponding to case 2 to take into account the 
impact of this information on the gain fractions. It can be seen that previous knowledge on the hierarchy plays a very important role for the performance of T2HK, while it is less significant for other new facilities (albeit, $\bar G$ is generically more affected than $G$). Color codes for each facility are shown in the legend.

Consider first Fig.~\ref{fig:gainfractions905sigdelta} (left), which shows $G$ as a function of the year when T2K+NO$\nu$A stop data taking.
On general grounds, if a positive hint of a CP-violating phase is found at the 90\% CL at T2K+NO$\nu$A the gain fraction will increase, since CP-violating values of $\delta$ become favored by data. 
When increasing the exposure, if we keep requiring only a 90\% CL hint for CP violation, the gain fraction will tend to a plateau. If we were to use real T2K+NO$\nu$A data instead of just requiring a 90\% CL hint, the CL at which CP violation would be favored should rather increase with the exposure (assuming that $\delta$ is indeed CP-violating) and hence the gain fraction would continue rising instead of reaching a plateau. In Fig.~\ref{fig:gainfractions905sigdelta} (left), when the mass hierarchy is known, the plateau is clearly visible for NF10 and T2HK. 
However, when the data from T2K+NO$\nu$A are complementary to that of the new facility and carry significant statistical weight, the gain fraction will continue to rise with the exposure of T2K+NO$\nu$A, since they play an important role in reaching the $5 \sigma$ goal. This behaviour is clearly seen for both LBNE setups, which show a monotonously increasing $G$ with T2K+NO$\nu$A running time, 
reaching $\sim 10\% (\sim 40\%)$ by 2030. For these facilities, the combination with more and more T2K+NO$\nu$A data will increase the overall probability to 
measure CP violation at 5$\sigma$. 

Notice also that the behavior of $G$ for NF10, LBNE-10 kton and LBNE-33 kton is practically unaffected by the previous knowledge of 
the neutrino mass hierarchy (represented by upper edges on the bands), and when the mass hierarchy is known the gain fraction $G$ only slightly increases. 
This is due to the fact that, at T2K+NO$\nu$A, positive (negative) CP-violating values of $\delta$ with a normal (inverted) hierarchy are very degenerate with 
CP-conserving values of $\delta$ and an inverted (normal) hierarchy. It can then happen that, for CP-conserving values of $\delta$, the CP-violating degenerate 
solution is actually favored at the 90\% CL from T2K+NO$\nu$A, thus leading to a fake hint. These cases will not lead to a discovery at any of these new facilities 
that have very good sensitivity to the mass hierarchy and will therefore be able to rule out these fake hints. Conversely, if the mass hierarchy was known, 
these fake hints would not take place at T2K+NO$\nu$A and a positive hint will more likely imply a truly CP-violating value for $\delta$, 
thus increasing the chance of discovery at the new facility, that is, the gain fraction.

A completely different behavior is shown by T2HK (blue band): in the case the mass hierarchy is not known we see that $G$ decreases with time, reaching a plateau 
at $G \sim 0.45$ by 2022. This behavior can be understood as follows:
if a CP-violating $\delta$ is detected for a very low exposure at T2K+NO$\nu$A, then $\delta$ is probably very near
to $\delta = - 90^\circ$ for a normal hierarchy or $\delta =  90^\circ$ for an inverted hierarchy, a value for which the combination of T2K+NO$\nu$A data with those of T2HK will have a very high chance to measure CP violation at 5$\sigma$. 
On the other hand, if CP violation is detected at 90\% CL for increasing exposure, different values of $\delta$ can give rise to the 90\% CL hint. 
For many of these values, the sign degeneracies will be localized at CP-conserving values. And, while a high exposure from T2K+NO$\nu$A can disfavor the CP-conserving degeneracies at the 90\% CL we are demanding, the combination of T2K+NO$\nu$A and T2HK cannot do the same at 5$\sigma$ and hence cannot discover CP violation. This explains the decreasing gain fraction of T2HK when increasing T2K+NO$\nu$A exposure if the mass hierarchy is not known. This problem is of course solved if the neutrino mass hierarchy were known when T2K+NO$\nu$A data are analized, 
as shown by the upper line of the T2HK blue band, for which we recover the expected increasing $G$ with time. Notice that also T2HK reaches
a plateau by 2020 (as the NF10), with $G \sim 0.7$. Running T2K+NO$\nu$A beyond this date, therefore, does not increase the discovery chances, 
unlike the case of the LBNE setups, which keep growing after 2020, albeit more slowly.

Let us consider now Fig.~\ref{fig:gainfractions905sigdelta} (right), showing $\bar G$ as a function of the T2K+NO$\nu$A total running time. 
On general grounds, when no hint for CP violation is found at T2K+NO$\nu$A as the exposure is increased, it is more and more likely that $\delta$ is close 
to a CP-conserving value, thus making a discovery of CP violation impossible. Therefore, we expect that
$\bar G$ will monotonously decrease with T2K+NO$\nu$A total running time. 
Notice that the knowledge of the neutrino mass hierarchy will have a negative impact on the $\bar G$ gain fractions (with the exception of T2HK), 
since CP-violating regions disfavored at the 90\% CL by T2K+NO$\nu$A with respect to CP-conserving ones with opposite hierarchy could actually turn out to be the true solution and lead to a $5 \sigma$ discovery at the new facility. These fake CP-conserving solutions cannot take place when the mass hierarchy is known and hence a negative hint from T2K+NO$\nu$A is more likely to imply a CP-conserving $\delta$ in this scenario, decreasing the value of $\bar{G}$.
In the case of T2HK, on the other hand, previous knowledge of the hierarchy would increase $\bar G$, since this facility is unable to measure
the hierarchy by its own at the required $5 \sigma$ level and, therefore, its performance is significantly affected by sign degeneracies. 
When we study $\bar G$ for T2HK, then, we notice that in case 1, represented by the lower edge of the blue band, we get $\bar G \sim 0.2 \, (0.05)$ for T2K+NO$\nu$A 
running until 2015 (2030). On the other hand in case 2, represented as for $G$ by the upper edge of the band, $\bar G$ goes from $\bar G \sim 0.45 \to 0.07$
in the same time interval. NF10 and LBNE-33 kton show the expected decreasing $\bar G$ with T2K+NO$\nu$A running time:  for NF10
we have $\bar G \sim  0.7 \to 0.4$ in case 1 and $\bar G \sim 0.7 \to 0.35$ in case 2, whereas for LBNE-33 kton we have $\bar G \sim 0.13 \to 0.10$ in case 1
and $\bar G \sim 0.13 \to 0$ in case 2. The case of LBNE-10 kton is a bit different:  the statistical power of this facility is limited to its capability of measuring the 
neutrino mass hierarchy and, therefore, $\bar G$ increases with time since the new facility increases the ability of T2K+NO$\nu$A actually making a discovery 
(by ruling out fake CP-conserving sign degeneracies). On the other hand, if previous knowledge of the hierarchy is at hand, the facility has no capability to discover CP violation by its own.

We can see that if T2K+NO$\nu$A see no hint of leptonic CP violation with increasing exposure, a clear gap between NF10 and the rest of facilities opens: 
we can fairly say that the super-beam--based experiments have a very low or null chance to measure CP violation, whereas NF10 would still have a 40\% chance
to measure it. On the other hand, if T2K+NO$\nu$A have a positive signal already, T2HK appears to be an interesting alternative to the neutrino factory, in particular if some other experiment is able to measure the neutrino mass hierarchy. 

\subsection{Measurement of the neutrino mass hierarchy}
\label{sec:resH}

\begin{figure}[htb!]
\begin{flushleft}
\begin{tabular}{cc}
	\includegraphics[width=7cm]{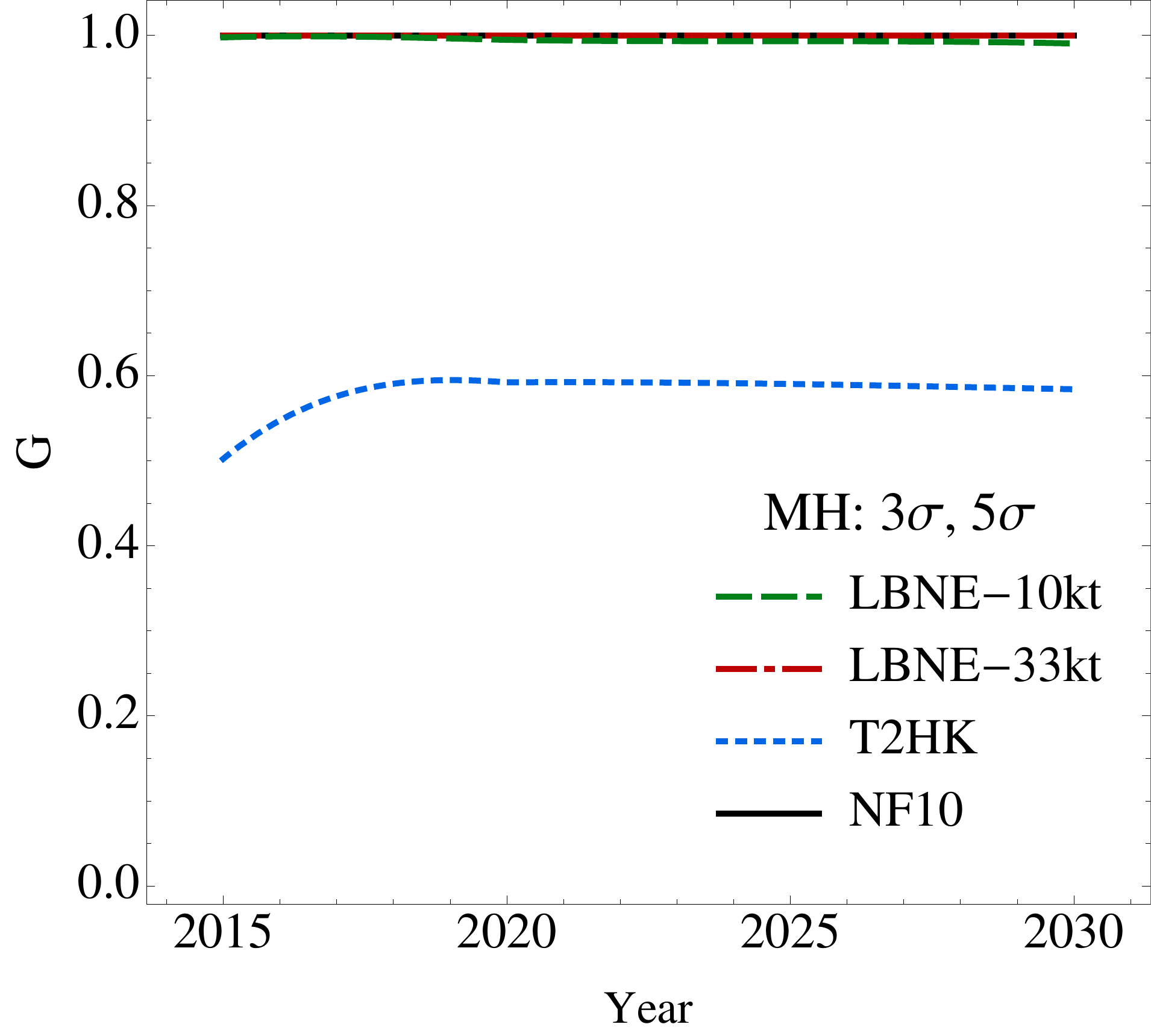} &
	\includegraphics[width=7cm]{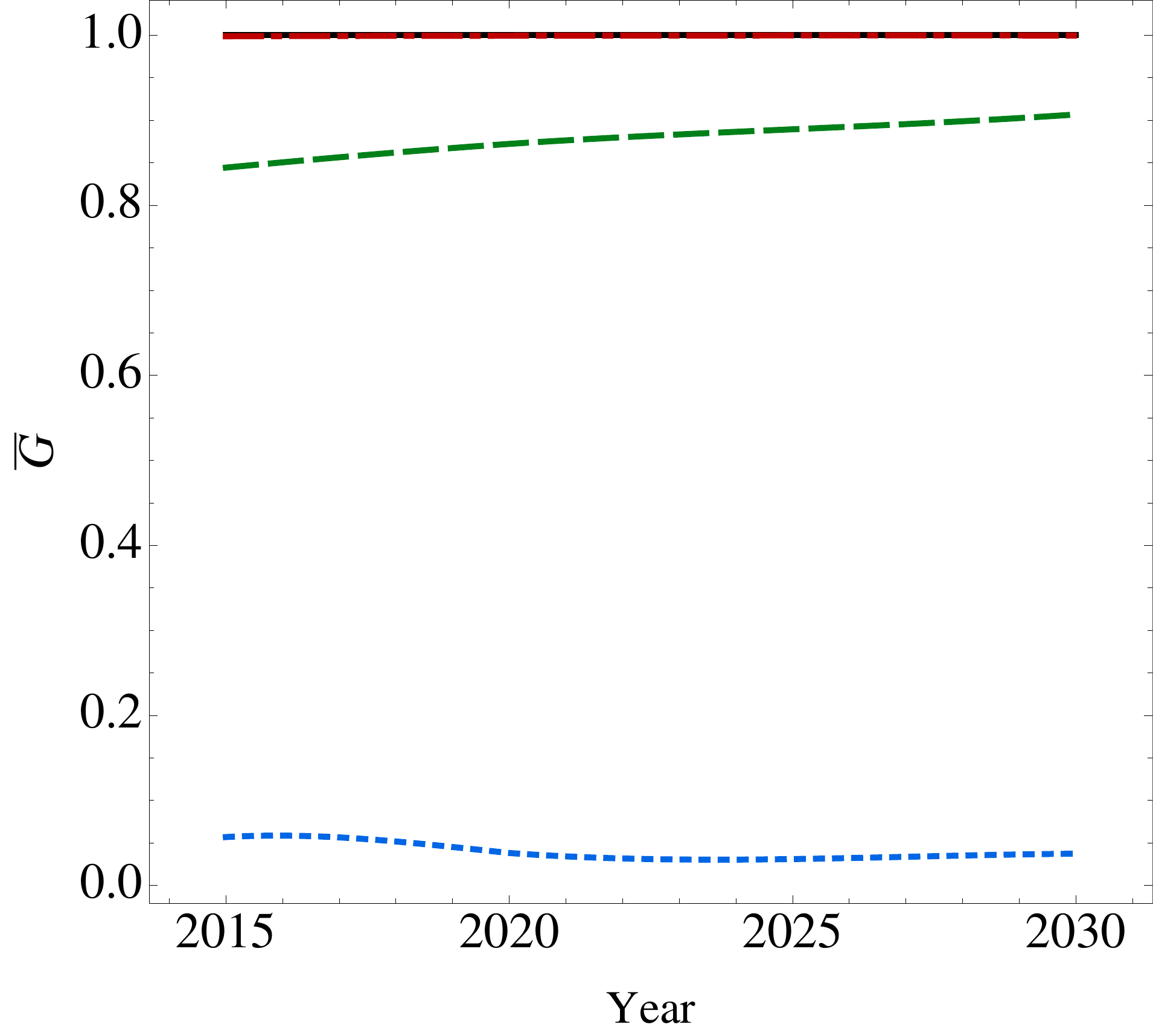}
\end{tabular}
\caption{\it The gain fractions $G$ (left panel) and $\bar G$ (right panel) for the mass hierarchy (MH)
defined as in Sect.~\ref{sec:method}, {\em i.e.} the probability that the new facility will identify a given hierarchy at
5$\sigma$ provided that T2K+NO$\nu$A have seen or not a hint of it at 90\%CL irrespectively of the value of $\delta$. In both cases results are shown as a function of  the year when T2K+NO$\nu$A stop taking data.
 }
\label{fig:gainfractions905sighierarchy}
\end{flushleft}
\end{figure}

Fig.~\ref{fig:gainfractions905sighierarchy} presents the gain fractions $G$ and $\bar G$ for the neutrino mass hierarchy discovery potential at 5$\sigma$ 
under the hypothesis of a positive (left panel) or negative (right panel) result at T2K+NO$\nu$A at $3 \sigma$, as a function of the T2K+NO$\nu$A total running time.
Consider first the gain fractions for NF10 and LBNE-33 kton. These two facilities have a very long baseline and sufficient statistics to measure the true neutrino mass hierarchy at 5$\sigma$ independently from the outcome of T2K+NO$\nu$A. We indeed see that $G  = \bar G = 1$ for both of them. 

On the other hand, for phase I of the LBNE experiment, LBNE-10 kton, we have $G \sim 1$ whereas $\bar G \sim 0.85$ depending on the outcome of T2K+NO$\nu$A. The reason for this is that both T2K+NO$\nu$A and LBNE-10 kton are more sensitive to the mass hierarchy for the same values of $\delta$. Therefore, if T2K+NO$\nu$A are not able to obtain a hint for the true mass hierarchy at $3\sigma$, then it is likely that $\delta$ lies in the small region of the parameter space where LBNE-10 kton cannot reach the $5\sigma$ level either.

Finally, the only facility whose gain fractions are strongly affected by the results of T2K+NO$\nu$A is T2HK, whose very short baseline makes it very difficult to 
measure the hierarchy by itself. We see in Fig.~\ref{fig:gainfractions905sighierarchy} (left) that $G$ increases with T2K+NO$\nu$A running time, reaching a plateau around  
$G \sim 0.6$ by 2017. Conversely, if T2K+NO$\nu$A are not able to find a $3 \sigma$ CL signal for the mass hierarchy with increasing exposure 
then the addition of T2HK will only allow a $5\sigma$ discovery with a $\sim 5\%$ probability.

\section{Conclusions}
\label{sec:concl}

In this paper we have introduced a new method to compare the physics reach of future neutrino oscillation facilities, taking into account the impact of the outcome of present long baseline experiments (T2K and NO$\nu$A).
We have compared the new facilities on the basis
of their capability to discover CP violation in the leptonic sector and to pin down the correct neutrino mass hierarchy, two of the missing inputs to complete
our phenomenological understanding of lepton mixing. 

In order to compare facilities, we have first computed the probability $P(B|\delta)$ that
T2K+NO$\nu$A find a 90\% CL hint for CP violation or a $3 \sigma$ indication for the mass hierarchy for a given $\delta$. 
We have done this by simulating 1000 possible outcomes of T2K+NO$\nu$A by randomly generating realizations of the experiment properly distributed
around the theoretical expected values for fixed leptonic mixing parameters (including $\delta$ and the neutrino mass hierarchy). 
We have then computed, applying the same method, the probability $P(A|\delta)$ with which one of the new facilities, in combination with T2K+NO$\nu$A, is able to 
discover CP violation or the neutrino mass hierarchy. Eventually, we have computed the {\em joint probability} $P(A,B|\delta)$ 
that T2K+NO$\nu$A provides a hint for CP violation (or measure the mass hierarchy) at a given CL and that the combined data from T2K+NO$\nu$A 
and the new facility make a discovery.
The {joint probability} $P(A,\bar B|\delta)$ is a way to measure the complementarity of the new facility with T2K+NO$\nu$A. Indeed, for the values of $\delta$ where this probability is large, there is a significant probability that CP violation will be missed by T2K+NO$\nu$A but still be discovered at the new facility.

After having evaluated the degree of complementarity between T2K+NO$\nu$A and the new facilities, we have introduced a very condensed performance
estimator, the {\em gain fraction} $G$. This is the {\em conditional probability} $P(A|B) = P(A,B)/P(B)$ that the new facility discovers CP violation 
(or the neutrino mass hierarchy) at 5$\sigma$ given that T2K+NO$\nu$A have been able ($G$) or unable ($\bar G$) to see a hint of it
at the 90\% CL ($3 \sigma$), regardless of the value of $\delta$. While the joint probability quantifies the complementarity between a new facility and T2K+NO$\nu$A, 
the {\em conditional probabilty} allows to make well-defined statistical statements on how the performance of a new facility will be affected by the different 
possible outcomes of T2K+NO$\nu$A. It is particularly interesting to see how $G$ and $\bar G$ for a given new facility scale with increasing 
T2K+NO$\nu$A exposure, {\em i.e.} with increasing T2K+NO$\nu$A running time. This is of extreme interest in a moment in which a decision on which facility
should be built must be taken, since it helps answering the questions: What would happen if we keep running T2K+NO$\nu$A beyond their scheduled data taking 
period? Is that helpful to improve the chances of new facilities to achieve their goals? Can the results from T2K+NO$\nu$A help to decide what is the optimal next step?

Armed with these concepts, we have compared three new facilities: T2HK, LBNE (in two versions, with a 10 kton and a 33 kton mass detector) and 
the 10 GeV Neutrino Factory (NF10). The outcome of our analysis shows that:
\begin{enumerate}
\item
NF10 is, as expected, the facility which shows the most improvement over T2K+NO$\nu$A. For example, the joint probability $P(A,\bar B|\delta)$ 
(\textit{i.e.}, the probability that T2K+NO$\nu$A do not find a 90\% CL hint for CP violation and that the combination with NF10 provides a 5$\sigma$ discovery) shows high peaks located at $\delta \sim \pm 30^\circ, \pm 150^\circ$, corresponding 
to regions of the $\delta$ parameter space that can be tested by NF10 but not by T2K+NO$\nu$A.
 On the other hand, peaks for T2HK (LBNE-33 kton) only appear for $\delta = -90^\circ$ ($\delta = 90^\circ$), {\em i.e.} at the same values for which T2K+NO$\nu$A 
 have the highest chances to discover CP violation. If we keep running T2K+NO$\nu$A until 2030, the joint probabilities for these two facilities vanish, showing 
 that the areas of the parameter space explored by T2HK and LBNE are less complementary to those probed by T2K+NO$\nu$A. 
 \item
In case of a positive hint for CP violation at T2K+NO$\nu$A at the 90\% CL, NF10 has a 
90\% chance to improve the statistical significance of the hint up to 5$\sigma$. The gain fraction reaches a plateau very early, showing that running T2K+NO$\nu$A
beyond, say, 2020 will not increase the gain fraction of NF10. The super-beam with highest chance to discover CP violation at 5$\sigma$ in case of 
a positive hint at T2K+NO$\nu$A is T2HK. In this case, $G$ also reaches a plateau for a rather low T2K+NO$\nu$A exposure, and running T2K+NO$\nu$A beyond
2020 does not increase their combined discovery potential significantly. The performance of T2HK, however, is strongly affected by our present ignorance of the neutrino mass hierarchy. If the hierarchy is still 
unknown at the time of T2HK data taking, then $G \sim 0.45$. On the other hand, if some other experiment is able to measure the hierarchy, we see
that $G$ increases significantly to $\sim 0.75$. Eventually, LBNE is the only facility whose gain fraction increases linearly with T2K+NO$\nu$A exposure. In this case, 
keeping both experiments running adds information to the new facility data, reaching $G \sim 0.45$ if T2K+NO$\nu$A keep running until 2030 ($G \sim 0.1$ when 
the 10~kton version is considered). 
\item
In case T2K+NO$\nu$A are not able to detect CP violation at the 90\% CL, all of the considered super-beams will have a very low discovery probability, 
even though for low T2K+NO$\nu$A exposure (say, until 2020) we can still see some hierarchy in the performances of T2HK and LBNE-33 kton (for which we get
$\bar G \sim 0.2$ and $\bar G \sim 0.1$, respectively). If we keep running T2K+NO$\nu$A until 2030 with still no hint of CP violation, then the chances
that one of these facilities will be able to discover CP violation at 5$\sigma$ will all be below 10\%. On the other hand, the gain fraction $\bar G$ at NF10 ranges
from $\bar G \sim 0.6$ in 2020 to $\bar G \sim 0.35$ in 2030. This is the only facility with a non-negligible probability to discover CP violation in the leptonic
sector if no hint of it is present even after running T2K+NO$\nu$A until 2030.
\item
The discovery of the true neutrino mass hierarchy at 5$\sigma$ is not a good observable to compare the performances of new facilities, as all of them (with the exception of T2HK) have been designed to maximize their probability to achieve this goal. However, it should be stated that if T2K+NO$\nu$A detect a 90\% CL hint on the true hierarchy by 2020, then the chances that T2HK can confirm such hint at 5$\sigma$ are as high as 60\% (all other facilities have $G = 1$ in this case). 
On the other hand, if T2K+NO$\nu$A have not seen a preferred hierarchy by 2020, then the chances that T2HK can detect the true hierarchy at 5$\sigma$
fall to less than 5\%. In this case, while NF10 and LBNE-33 kton still show $\bar G = 1$, the smaller version of LBNE will have a very good chance to discover
the neutrino mass hierarchy but it will not be guaranteed (with $\bar G$ ranging from 0.85 to 0.90 if T2K+NO$\nu$A are kept running until 2020 or 2030, respectively).
\end{enumerate}

\section*{Acknowledgements}

We acknowledge very useful discussions with P.~Hernandez. 
We also acknowledge partial support by European Union FP7 ITN INVISIBLES (Marie
Curie Actions, PITN-GA-2011-289442); the Spanish MINECO 
through the project FPA2009-09017, the Ram\'on y Cajal programme (RYC-2011-07710)  
and the Consolider-Ingenio CUP (CSD2008-00037); the Comunidad Aut\'onoma de Madrid
through the project HEPHACOS P-ESP-00346; and by the U.S. Department of Energy under award number \protect{DE-SC0003915}.


\begin{thebibliography}{10}

\bibitem{Pontecorvo:1957cp}
B. Pontecorvo,
\newblock Sov.Phys.JETP 6 (1957) 429,
\newblock 

\bibitem{Pontecorvo:1957qd}
B. Pontecorvo,
\newblock Sov.Phys.JETP 7 (1958) 172,
\newblock 

\bibitem{Maki:1960ut}
Z. Maki, M. Nakagawa, Y. Ohnuki and S. Sakata,
\newblock Prog.Theor.Phys. 23 (1960) 1174,
\newblock 

\bibitem{Maki:1962mu}
Z. Maki, M. Nakagawa and S. Sakata,
\newblock Prog.Theor.Phys. 28 (1962) 870,
\newblock 

\bibitem{Pontecorvo:1967fh}
B. Pontecorvo,
\newblock Sov.Phys.JETP 26 (1968) 984,
\newblock 

\bibitem{Cleveland:1998nv}
B. Cleveland et~al.,
\newblock Astrophys.J. 496 (1998) 505,
\newblock 

\bibitem{Kaether:2010ag}
F. Kaether, W. Hampel, G. Heusser, J. Kiko and T. Kirsten,
\newblock Phys.Lett. B685 (2010) 47, 1001.2731,
\newblock 

\bibitem{Abdurashitov:2009tn}
SAGE Collaboration, J. Abdurashitov et~al.,
\newblock Phys.Rev. C80 (2009) 015807, 0901.2200,
\newblock 

\bibitem{Hosaka:2005um}
Super-Kamiokande Collaboration, J. Hosaka et~al.,
\newblock Phys.Rev. D73 (2006) 112001, hep-ex/0508053,
\newblock 

\bibitem{Aharmim:2007nv}
SNO Collaboration, B. Aharmim et~al.,
\newblock Phys.Rev. C75 (2007) 045502, nucl-ex/0610020,
\newblock 

\bibitem{Aharmim:2005gt}
SNO Collaboration, B. Aharmim et~al.,
\newblock Phys.Rev. C72 (2005) 055502, nucl-ex/0502021,
\newblock 

\bibitem{Aharmim:2008kc}
SNO Collaboration, B. Aharmim et~al.,
\newblock Phys.Rev.Lett. 101 (2008) 111301, 0806.0989,
\newblock 

\bibitem{Bellini:2011rx}
The Borexino Collaboration, G. Bellini et~al.,
\newblock Phys.Rev.Lett. 107 (2011) 141302, 1104.1816,
\newblock 

\bibitem{An:2012eh}
DAYA-BAY Collaboration, F. An et~al.,
\newblock Phys.Rev.Lett. 108 (2012) 171803, 1203.1669,
\newblock 

\bibitem{Ahn:2012nd}
RENO collaboration, J. Ahn et~al.,
\newblock Phys.Rev.Lett. 108 (2012) 191802, 1204.0626,
\newblock 

\bibitem{Abe:2011fz}
DOUBLE-CHOOZ Collaboration, Y. Abe et~al.,
\newblock Phys.Rev.Lett. 108 (2012) 131801, 1112.6353,
\newblock 

\bibitem{Wendell:2010md}
Super-Kamiokande Collaboration, R. Wendell et~al.,
\newblock Phys.Rev. D81 (2010) 092004, 1002.3471,
\newblock 

\bibitem{Gando:2010aa}
KamLAND Collaboration, A. Gando et~al.,
\newblock Phys.Rev. D83 (2011) 052002, 1009.4771,
\newblock 

\bibitem{Ahn:2002up} 
K2K Collaboration, M. H. Ahn et~al., 
\newblock  Phys.Rev.Lett. 90 (2003) 041801, hep-ex/0212007,
\newblock 

\bibitem{Ahn:2006zza}
K2K Collaboration, M. Ahn et~al.,
\newblock Phys.Rev. D74 (2006) 072003, hep-ex/0606032,
\newblock 

\bibitem{Adamson:2011qu}
MINOS Collaboration, P. Adamson et~al.,
\newblock Phys.Rev.Lett. 107 (2011) 181802, 1108.0015,
\newblock 

\bibitem{Abe:2011sj}
T2K Collaboration, K. Abe et~al.,
\newblock Phys.Rev.Lett. 107 (2011) 041801, 1106.2822,
\newblock 

\bibitem{Adamson:2012rm}
MINOS Collaboration, P. Adamson et~al.,
\newblock Phys.Rev.Lett. 108 (2012) 191801, 1202.2772,
\newblock 

\bibitem{Abe:2012gx}
T2K Collaboration, K. Abe et~al.,
\newblock Phys.Rev. D85 (2012) 031103, 1201.1386,
\newblock 

\bibitem{GonzalezGarcia:2012sz}
M. Gonzalez-Garcia, M. Maltoni, J. Salvado and T. Schwetz,
\newblock (2012), 1209.3023,
\newblock 

\bibitem{Tortola:2012te}
D. Forero, M. Tortola and J. Valle,
\newblock Phys.Rev. D86 (2012) 073012, 1205.4018,
\newblock 

\bibitem{Fogli:2012ua}
G. Fogli et~al.,
\newblock Phys.Rev. D86 (2012) 013012, 1205.5254,
\newblock 

\bibitem{Beringer:1900zz}
Particle Data Group, J. Beringer et~al.,
\newblock Phys.Rev. D86 (2012) 010001,
\newblock 

\bibitem{DescotesGenon:2012rq}
S. Descotes-Genon,
\newblock (2012), 1209.4016,
\newblock 

\bibitem{Ricciardi:2013mca}
S. Ricciardi,
\newblock (2013), 1302.4582,
\newblock 

\bibitem{Itow:2001ee}
T2K Collaboration, Y. Itow et~al.,
\newblock (2001) 239, hep-ex/0106019,
\newblock 

\bibitem{Ayres:2004js}
NOvA Collaboration, D. Ayres et~al.,
\newblock (2004), hep-ex/0503053,
\newblock 

\bibitem{Prakash:2012az}
S. Prakash, S.K. Raut and S.U. Sankar,
\newblock Phys.Rev. D86 (2012) 033012, 1201.6485,
\newblock 

\bibitem{novawhitepaper}
M. Messier for~the NOvA~Collaboration,
\newblock Whitepaper submitted to SNOWMASS, 2013.

\bibitem{Blennow:2012gj}
M. Blennow and T. Schwetz,
\newblock JHEP 1208 (2012) 058, 1203.3388,
\newblock 

\bibitem{Akhmedov:2012ah}
E.K. Akhmedov, S. Razzaque and A.Y. Smirnov,
\newblock (2012), 1205.7071,
\newblock 

\bibitem{Ghosh:2012px}
A. Ghosh, T. Thakore and S. Choubey,
\newblock (2012), 1212.1305,
\newblock 

\bibitem{Agarwalla:2012uj}
S.K. Agarwalla, T. Li, O. Mena and S. Palomares-Ruiz,
\newblock (2012), 1212.2238,
\newblock 

\bibitem{Franco:2013in}
D. Franco et~al.,
\newblock (2013), 1301.4332,
\newblock 

\bibitem{Petcov:2001sy}
S. Petcov and M. Piai,
\newblock Phys.Lett. B533 (2002) 94, hep-ph/0112074,
\newblock 

\bibitem{Choubey:2003qx}
S. Choubey, S. Petcov and M. Piai,
\newblock Phys.Rev. D68 (2003) 113006, hep-ph/0306017,
\newblock 

\bibitem{Zhan:2008id}
L. Zhan, Y. Wang, J. Cao and L. Wen,
\newblock Phys.Rev. D78 (2008) 111103, 0807.3203,
\newblock 

\bibitem{Zhan:2009rs}
L. Zhan, Y. Wang, J. Cao and L. Wen,
\newblock Phys.Rev. D79 (2009) 073007, 0901.2976,
\newblock 

\bibitem{Qian:2012xh}
X. Qian et~al.,
\newblock (2012), 1208.1551,
\newblock 

\bibitem{Qian:2012zn}
X. Qian et~al.,
\newblock (2012), 1210.3651,
\newblock 

\bibitem{Ciuffoli:2012bp}
E. Ciuffoli et~al.,
\newblock (2012), 1211.6818,
\newblock 

\bibitem{Ciuffoli:2012bs}
E. Ciuffoli, J. Evslin and X. Zhang,
\newblock JHEP 1212 (2012) 004, 1209.2227,
\newblock 

\bibitem{Blennow:2009pk}
M. Blennow and E. Fernandez-Martinez,
\newblock Comput.Phys.Commun. 181 (2010) 227, 0903.3985,
\newblock 

\bibitem{Schwetz:2006md}
T. Schwetz,
\newblock Phys.Lett. B648 (2007) 54, hep-ph/0612223,
\newblock 

\bibitem{Patterson:2012zs}
NOvA Collaboration, R. Patterson,
\newblock (2012), 1209.0716,
\newblock 

\bibitem{T2Kexpo}
T. Koseki,
\newblock talk at Open meeting for the Hyper-Kamiokande Project, August 2012.

\bibitem{Huber:2009cw}
P. Huber, M. Lindner, T. Schwetz and W. Winter,
\newblock JHEP 0911 (2009) 044, 0907.1896,
\newblock 

\bibitem{Agarwalla:2012bv}
S.K. Agarwalla, S. Prakash, S.K. Raut and S.U. Sankar,
\newblock (2012), 1208.3644,
\newblock 

\bibitem{Geer:2007kn}
S. Geer, O. Mena and S. Pascoli,
\newblock Phys.Rev. D75 (2007) 093001, hep-ph/0701258,
\newblock 

\bibitem{Ankenbrandt:2009zza}
C. Ankenbrandt et~al.,
\newblock Phys.Rev.ST Accel.Beams 12 (2009) 070101,
\newblock 

\bibitem{FernandezMartinez:2010zza}
E. Fernandez~Martinez, T. Li, S. Pascoli and O. Mena,
\newblock Phys.Rev. D81 (2010) 073010, 0911.3776,
\newblock 

\bibitem{Dighe:2011pa}
A. Dighe, S. Goswami and S. Ray,
\newblock Phys.Rev. D86 (2012) 073001, 1110.3289,
\newblock 

\bibitem{Ballett:2012rz}
P. Ballett and S. Pascoli,
\newblock Phys.Rev. D86 (2012) 053002, 1201.6299,
\newblock 

\bibitem{Abe:2011ts}
K. Abe et~al.,
\newblock (2011), 1109.3262,
\newblock 

\bibitem{HKnuturn}
M. Shiozawa,
\newblock talk at NuTURN 2012, May 2012.

\bibitem{openHKmeeting}
M. Yokoyama,
\newblock talk at the First open meeting for the Hyper-Kamiokande project,
  August 2012.

\bibitem{Campagne:2006yx}
J.E. Campagne, M. Maltoni, M. Mezzetto and T. Schwetz,
\newblock JHEP 0704 (2007) 003, hep-ph/0603172,
\newblock 

\bibitem{Coloma:2011pg}
P. Coloma and E. Fernandez-Martinez,
\newblock JHEP 1204 (2012) 089, 1110.4583,
\newblock 

\bibitem{Baussan:2012cw}
E. Baussan et~al.,
\newblock (2012), 1212.5048,
\newblock 

\bibitem{LBNO}
A. Stahl et~al.,
\newblock (2012),
\newblock 

\bibitem{Akiri:2011dv}
LBNE Collaboration, T. Akiri et~al.,
\newblock (2011), 1110.6249,
\newblock 

\bibitem{Choubey:2011zzq}
IDS-NF Collaboration, S. Choubey et~al.,
\newblock (2011), 1112.2853,
\newblock 

\bibitem{Agarwalla:2010hk}
S.K. Agarwalla, P. Huber, J. Tang and W. Winter,
\newblock JHEP 1101 (2011) 120, 1012.1872,
\newblock 

\bibitem{Bayes:2012ex}
R. Bayes et~al.,
\newblock Phys.Rev. D86 (2012) 093015, 1208.2735,
\newblock 

\bibitem{bayes}
R. Bayes,
\newblock Private communication.

\bibitem{Indumathi:2009hg}
D. Indumathi and N. Sinha,
\newblock Phys.Rev. D80 (2009) 113012, 0910.2020,
\newblock 

\bibitem{Donini:2010xk}
A. Donini, J. Gomez~Cadenas and D. Meloni,
\newblock JHEP 1102 (2011) 095, 1005.2275,
\newblock 

\bibitem{Coloma:2011zza}
P. Coloma,
\newblock AIP Conf.Proc. 1382 (2011) 121,
\newblock 

\bibitem{Dutta:2011mc}
R. Dutta, D. Indumathi and N. Sinha,
\newblock Phys.Rev. D85 (2012) 013003, 1103.5578,
\newblock 

\bibitem{Coloma:2012ji}
P. Coloma, P. Huber, J. Kopp and W. Winter,
\newblock (2012), 1209.5973,
\newblock 

\bibitem{Martini:2012fa}
M. Martini, M. Ericson and G. Chanfray,
\newblock Phys.Rev. D85 (2012) 093012, 1202.4745,
\newblock 

\bibitem{Nieves:2012yz}
J. Nieves, F. Sanchez, I. Ruiz~Simo and M. Vicente~Vacas,
\newblock Phys.Rev. D85 (2012) 113008, 1204.5404,
\newblock 

\bibitem{Lalakulich:2012hs}
O. Lalakulich and U. Mosel,
\newblock Phys.Rev. C86 (2012) 054606, 1208.3678,
\newblock 

\bibitem{Minakata:2001qm}
H. Minakata and H. Nunokawa,
\newblock JHEP 0110 (2001) 001, hep-ph/0108085,
\newblock 

\bibitem{Donini:2003vz}
A. Donini, D. Meloni and S. Rigolin,
\newblock JHEP 0406 (2004) 011, hep-ph/0312072,
\newblock 

\bibitem{Huber:2002mx}
P. Huber, M. Lindner and W. Winter,
\newblock Nucl.Phys. B645 (2002) 3, hep-ph/0204352,
\newblock 

\end{thebibliography}

%
\end{document}